\makeatletter
\providecommand*{\input@path}{}
\g@addto@macro\input@path{{style/}}
\makeatother

\documentclass{article}
\usepackage[T1]{fontenc}
\usepackage[utf8]{inputenc}
\usepackage{ismir, amsmath, amssymb, cite, url, graphicx, color}
\def\lbd{}	

\title{Musical Audio Similarity with Self-supervised Convolutional Neural Networks}

\multauthor{Carl Thomé$^1$ \hspace{1cm} Sebastian Piwell$^1$ \hspace{1cm} Oscar Utterbäck$^1$}{
\\$^1$ Epidemic Sound, Stockholm, Sweden\\
{\tt\small firstname.lastname@epidemicsound.com}
}

\def\authorname{C. Thomé, S. Piwell, and O. Utterbäck}

\usepackage[bookmarks=false,pdfauthor={\authorname},pdfsubject={\papersubject},hidelinks]{hyperref}
\begin{document}
\maketitle

\begin{abstract}
We have built a music similarity search engine that lets video producers search by listenable music excerpts, as a complement to traditional full-text search. Our system suggests similar sounding track segments in a large music catalog by training a self-supervised convolutional neural network with triplet loss terms and musical transformations. Semi-structured user interviews demonstrate that we can successfully impress professional video producers with the quality of the search experience, and perceived similarities to query tracks averaged 7.8/10 in user testing. We believe this search tool will make for a more natural search experience that is easier to find music to soundtrack videos with.
\end{abstract}

\section{Introduction}\label{sec:introduction}
Finding songs to soundtrack with is crucial for video production. Rather than browsing by track metadata (genres, moods, musical key, instrumentation), we believe browsing by example is an intuitive alternative as the user does not need to provide textual descriptions \cite{grosche2012audio}. Instead, by providing example songs and how they \emph{sound}, we want to let video producers find similar tracks in a music catalog.

By having an algorithmic notion of what constitutes music similarity for video production, we could simply measure pairwise distances from a query audio and recommend neighboring catalog tracks. However, it is difficult to design such algorithms for a general setting\cite{schluter2011music, pohle2009rhythm}: the search can become overly strict, retrieving poor matches that break the rules, or entirely miss good matches not covered by the rules. Thus, many systems rely on learning similarities by an objective function and example data\cite{schluter2011music, slaney2008learning, dieleman2012learning} to relax assumptions before the retrieval task.

In this work, we've trained similarity learning models by triplet loss terms, with the ambition of finding a distance measure where similar sounding clips should be close and dissimilar sounding clips should be distant\cite{wu2017sampling, jansen2018unsupervised}. In order to design our notion of similarity, we use music production knowledge to design rules for what constitutes a positive and negative example in the triplet loss formulation.

Our contribution is empirical evidence that self-supervised similarity learning can be valuable for music similarity search in a large scale production environment in an industry setting.

\section{Design}\label{sec:method}

\begin{figure}
    \centering
    \includegraphics[width=0.9\columnwidth]{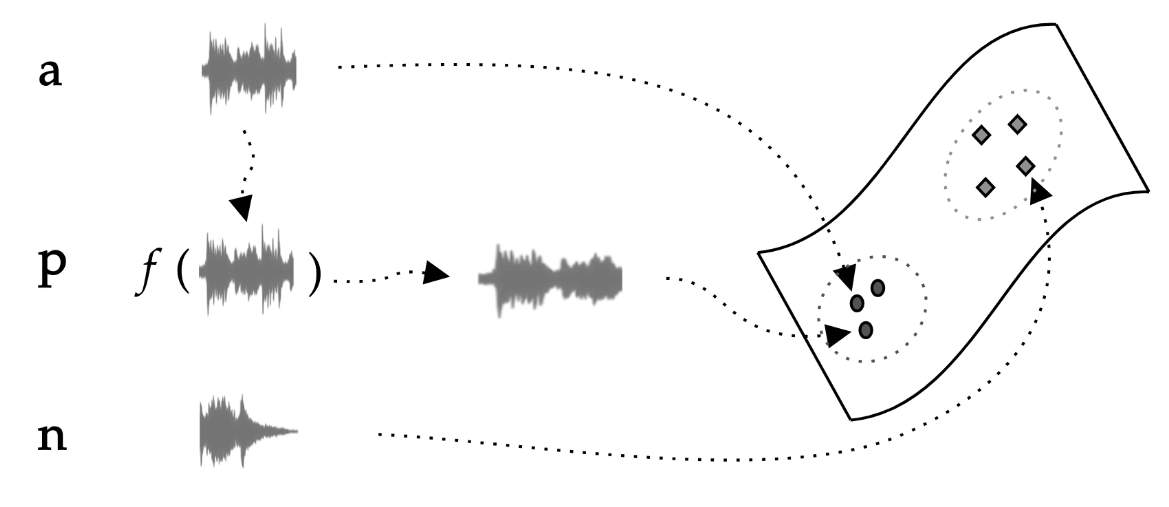}
    \caption{An overview of our self-supervised similarity learning approach for musical audio. We construct training data by transforming audio clips with a randomized audio effects chain of musical transformations.}
    \label{fig:overview}
\end{figure}

In order to retrieve similar sounding tracks from a given query song, we need to find parameters $\theta$ for an encoding function $e_\theta$ such that pairwise distances between query songs and tracks in the catalog will agree with video producers, meaning $$d(e_\theta(\mathbf{a}), e_\theta(\mathbf{p})) < d(e_\theta(\mathbf{a}), e_\theta(\mathbf{n}))$$ where $d$ is Euclidean distance, and the anchor $\mathbf{a}$ is any query song. $\mathbf{n}$ is a negative example song not deemed similar to the anchor, and $\mathbf{p}$ is a positive example song perceived as similar to the anchor. However, we lack annotated triplets $(\mathbf{a}, \mathbf{n}, \mathbf{p})$ from video producers, so we've designed an audio effects chain of musical transformations $f$ that we believe $e_\theta$ should be approximately invariant for (see \figref{fig:overview}), as in $$e_\theta(\mathbf{x}_0) \approx e_\theta(f(\mathbf{x}_0)).$$

During training the audio effects are applied stochastically by turning them on and off and tweaking their settings per audio clip, and the goal is to steer the learned similarity by including this domain knowledge.\footnote{This is a powerful benefit of our approach as future developers can tune the system for changing requirements without having to understand historical annotators, since the similarity notion is explicit in code.}

\subsection{Data}\label{sec:data}
From our music catalog of 40,000 tracks, we randomly assign 80\% of tracks to training, 10\% for validation and 10\% for testing by hashing track ids. We use validation tracks for model development (learning rate annealing, early stopping and model selection) and testing tracks exclusively for estimating generalization upon deployment. We downsample and sum each full mix track to monophonic 16 kHz with librosa \cite{librosa}, and sample ten random clips at 10.0 seconds each from every track with Dataflow \cite{akidau2015dataflow}, such that long tracks don't dominate in model training.

\subsection{Encoder}\label{sec:encoder}
The encoder structure consists of a spectral transform from audio signals to Mel spectrograms, with online feature standardization by batch normalization, and a standard ConvNet for embedding computation. The spectral transform has a DFT size of 2048 samples, windowed by 50\% overlapping Hann windows. The Mel bands are triangular filters starting from 20 Hz up to the Nyquist limit as provided in TensorFlow \cite{tensorflow}. Magnitudes are log scaled to promote timbral features and not only pitch and rhythm. To project into RGB space, an initial 2D convolution runs between the spectrogram and the ConvNet. We view the ConvNet variant itself as a categorical hyperparameter during model development but the default classification head is replaced with a fully-connected layer with 128 output units and a unit normalizing activation function such that embeddings are on the surface of a hypersphere in $\mathbb{R}^{128}$ with radius one.

\subsection{Objective}\label{sec:objective}
The most important puzzle piece in our system is how the encoder receives parameters $\theta$. We rely on an objective of competing triplet loss terms.

For every input clip, we transform it into a positive by applying the stochastic audio effects chain $f$, consisting of a random time shift, time stretch, pitch shift, reverb and additive noise from a truncated Gaussian. The time stretching and pitch shifting is implemented as a phase vocoder while the reverb is implemented as multiplication in the frequency domain with exponentially decaying white noise.

We perform online triplet mining\cite{schroff2015facenet} and deem audio clips within a minibatch as positives if they share track metadata. We weigh triplet loss terms for transformed clips, clips from the same track, and genre and mood membership, with weights 1.0, 0.5, 0.1, 0.1 respectively. Model parameter and gradient estimate updates are computed with Adam\cite{kingma2014adam} for minibatches of 256 audio clips.

\section{Evaluation}
To know that encoders produce meaningful points in the embedding space such that neighboring tracks are perceived as similar, we conducted qualitative tests. However, for choosing models to send out for testing, we relied on offline metrics in \cite{jansen2018unsupervised} and computed average precision per category (one per genre, one per mood, etc.) on a ranked list of pairwise distances between embeddings where we put a one if both embeddings belong to the same category, and a zero if they belong to different categories.

\subsection{Quantitative}
Intuitively we want a clear separation between embedded audio clips from vastly different musical genres and moods. Similarly, we likely want clips from the same track to cluster together.\footnote{However, if we get perfect scores we arguably haven't accomplished anything meaningful, as we then could simply resort to using class memberships directly to recommend tracks.} 

We have computed mean average precision (mAP) as in \cite{jansen2018unsupervised} in \tabref{tab:test_scores} for our encoder, a random encoder, and a baseline encoder. The random encoder draws an embedding from a uniform distribution per clip and uses that as its representation. The baseline encoder computes 20 MFCC coefficients from 128 band Mel spectrograms and decorrelates the MFCC dimensions with incremental PCA with librosa and scikit-learn \cite{librosa, sklearn}.

\begin{table}[h]
    \begin{center}
    \begin{tabular}{|l|c|c|c|}
    \hline
                    & Genre             & Mood              & Track \\
    \hline
    Random          & 0.3143            & 0.3320            & 0.0495 \\
    Baseline        & 0.4011            & 0.3684            & 0.6054 \\
    ConvNet         & \textbf{0.7095}   & \textbf{0.5087}   & \textbf{0.8490} \\
    \hline
    \end{tabular}
    \end{center}
    \caption{Mean average precision (mAP) scores on test tracks for different sets of annotations and encoders.} \label{tab:test_scores}
\end{table}

\subsection{Qualitative}
Acknowledging that music similarity is hard to evaluate without humans\cite{dieleman2012learning}, we conducted semi-structured interviews with five professional video producers in which they explored a production grade web app for music discovery and were tasked with finding similar music from three starting tracks of their own choosing.

The mean opinion score when asked how similar they perceived the search results to be to their chosen query tracks was 7.8/10. The interviews were recorded and transcribed and can be shared upon request.

\section{Discussion}
After seeing video producers interact with the music similarity search engine, we believe self-supervised similarity learning on musical audio makes for a natural search experience to find music to soundtrack videos with. By circumventing annotation needs and not requiring users to express what they want beyond providing listenable examples, we can help video producers find suitable tracks easier and faster.

In future work, we will explore stereophonic signals, multi-track recordings, alternative loss functions\cite{saeed2021contrastive} and musical transformations\cite{spijkervet2021contrastive}, and ways of granting user control to the similarity measure\cite{lee2020disentangled}.


\clearpage

\bibliography{references}

\end{document}